# The Pierre Auger Cosmic Ray Observatory – Comments on Recent Results

Carlos Hojvat (for the Pierre Auger Collaboration)
*Fermilab, P.O. Box 500, Batavia, IL 60510-0500, USA*

The Pierre Auger Observatory has been taking data on the highest energy cosmic rays for the equivalent of over a year of full detector aperture. Comments are presented on results published so far.

## 1. INTRODUCTION

Data taking at the Pierre Auger Observatory has been proceeding since 2004, in parallel with the completion of the observatory. At the end of 2007 available data was equivalent to a full year of data taking with the completed Observatory. Although the rate of the highest energy events is the highest of all previous cosmic rays experiments, they are still few in numbers to fully understand the details of the detector. Further improvements in understanding the energy calibration and the detector's systematic are expected. The full published data will not be reviewed here [1-7]. We will comment on: if the energy calibration is known; are we actually seeing the GZK cut-off; if we see a correlation with probable sources only in the Northern Hemisphere.

## 2. THE AUGER HYBRID DETECTOR

The Pierre Auger Observatory is located north east of the town of Malargüe, Province of Mendoza, Argentina [9]. It covers 3000 km$^2$ with 1663 surface detectors (SD). The atmosphere is seen by 24 fluorescence telescopes (FD), located in 4 buildings at slightly elevated sites, Figure 1. The Observatory combines the two detection techniques, SD and FD. The SD provides the transverse distribution of showers at the surface. The FD provides the longitudinal development of the showers in the atmosphere, can only operate in dark nights or approximately 10% of the observation time.

## 3. RESULTS

### 3.1. Energy Calibration

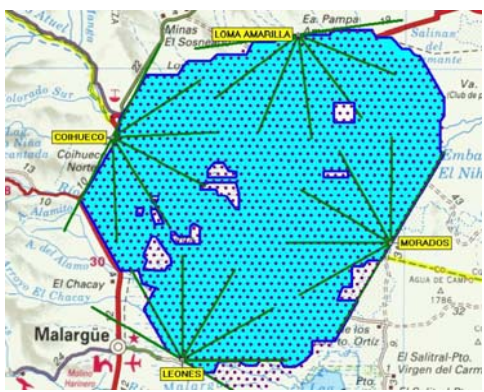
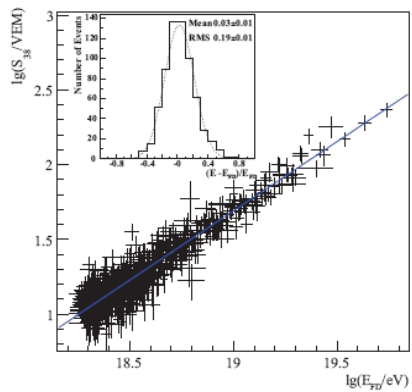

Figure 1: The surface and fluorescence detectors of the Pierre Auger Observatory. East-West dimension is 70 Km

Figure 2: Energy deposited in the SD vs. the reconstructed FD Energy [1]

The energy of events recorded by the SD is expressed in terms of the energy that would be deposited by a vertical muon (VEM). Energy calibration of the SD energy can be obtained by simulation. The energy deposited in the atmosphere by the shower is measured by the FD and can be directly related to the cosmic ray energy.

The energy of the SD events is cross calibrated between the SD VEM and the FD measurement. The calibration statistical accuracy is given by the number of "hybrid" events, those detected by both SD and FD detectors. This is shown in Figure 2, although a large number of hybrid events have been recorded, the correlation at the highest energies is given by a small number of events, a power law relationship is assumed.

The accuracy of the FD energy measurement is traced back to other absolute measurements [1]. Most important is the cross sections for producing fluorescent light in the atmosphere. Measurements of the atmosphere transparency, the atmosphere vertical profile and day/night differences on the detector are important.

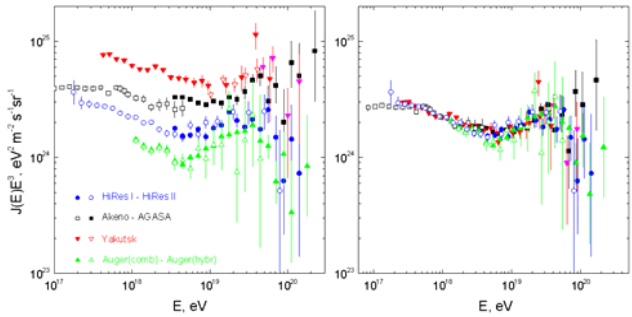
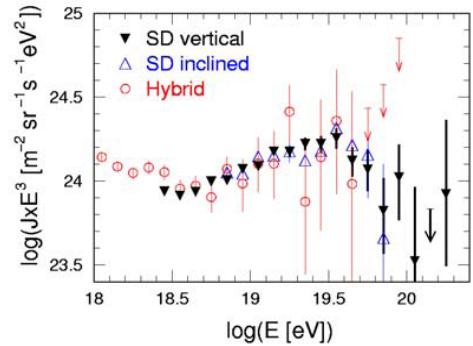

Figure 3: Comparison of the energy spectra of the highest energy cosmic rays from different experiments [10].

Figure 4: Auger spectra measured with vertical, inclined and hybrid events [11].

Comparison between spectra measured by different cosmic rays experiments have been published [10], Figure 3. To bring the different energy spectra in agreement, scaling has been applied: Auger (x1.2); Agasa (x0.75); Akeno (x0.83) and Yakutsk (x0.625), Figure 3 right panel. The 20% difference between Auger and HiRes is well within the quoted errors.

An internal check on the spectrum was performed by comparing different sets of Auger events, Figure 4 [11]. There is good agreement between spectra obtained with vertical, inclined (60 to 80 degrees of azimuth) and hybrid events. Figure 5 shows a comparison with HiRes results [12].

### 3.2. Energy Spectrum and flux suppression at the highest energies

The energy spectrum measured by the Auger Observatory is shown in Figure 6 [1]. The suppression of flux at the highest energy observed by both experiments, HiRes [12] and Auger [1] is consistent with the GZK effect [13]. The actual "shape" of the spectrum can depend also on the injection spectrum at the source, the primary composition, the number and types of sources, and the distance to them. If the observed flux suppression is given solely by the GZK mechanism will become clear as the number of events recorded at the highest energies increases.

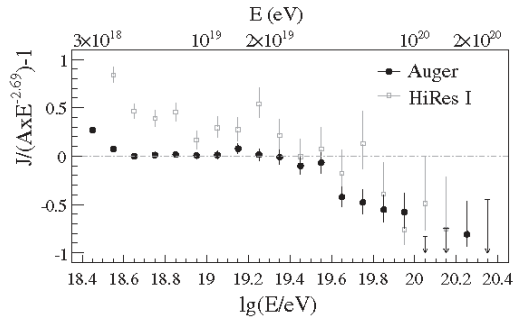 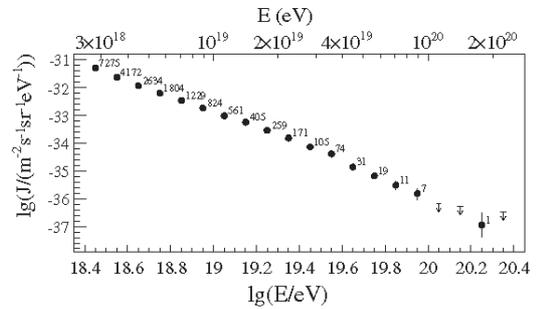

Figure 5: Comparison at the highest energies Auger events with the HiRes data [12]

Figure 6: The cosmic ray spectrum measured by the Auger Observatory [1].

### 3.3. Correlation with sources.

The Auger Observatory has published a correlation of 27 of the highest energy events with Active Galactic Nuclei [14], as seen from the Southern Hemisphere. Data indicates an accumulation of events from the region of Centaurus A and no events from the Virgo Cluster. The data is shown in Figure 7, an Aitoff projection of the celestial sphere in galactic coordinates. Circles of radius 3.1° are centered at the arrival directions. The positions of the 472 AGN with redshift $z$ < 0.018 from the Véron-Cetty catalog of quasars and active nuclei [15] are indicated by red asterisks.

A similar analysis was performed by the HiRes collaboration [16] with data taken at the Northern Hemisphere indicating no correlation with AGN, Figure 8. The black dots are the locations of AGN and 14 QSOs with redshift z < 0.018. The green circle and triangle mark the locations of Centaurus A and M87, respectively. Red circles (with radii of 3.1◦) mark 2 correlated events. Blue squares mark the locations of the 11 uncorrelated events.. Note the opposite meaning of the shaded areas in the two figures.

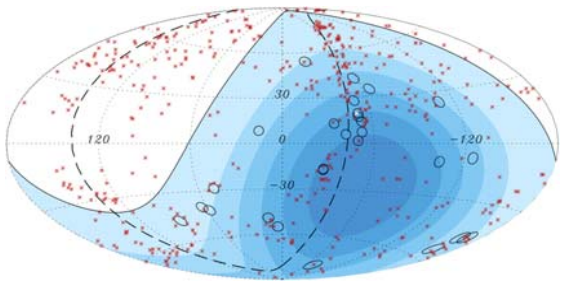 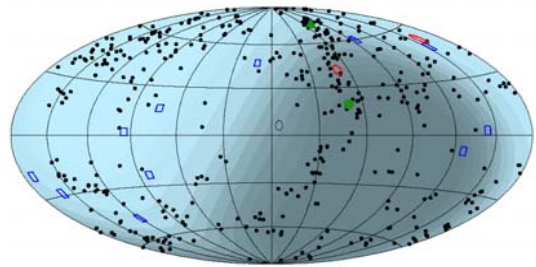

Figure 7: Auger Observatory- The solid line represents the border of the field of view. Darker color indicates larger relative exposure.

Figure 8: HiRes - The blue shaded regions delineate areas of even exposure in HiRes (lighter shades of blue indicate a greater exposure).

The two data sets seem only to agree on the lack of events from the Virgo cluster. Ignoring possible large differences in the nature of cosmic ray sources on the Northern and the Southern Hemispheres, the two sets of data are inconsistent. The Auger events, E>56 EeV, are at the edge of the flux suppression, different energy scales could produce differences on the number of selected events. The HiRes aperture has been estimated at 2400 Km² sr yr [17], for 13 events at E > 56

EeV. The Auger data correspond to a total aperture of 9000 Km2 sr yr for 27 events, for the same energy threshold. Based on the Auger rate we would expect only around 7 events in HiRes, half the ones indicated. One possible explanation for the disagreement is that the energy of the HiRes events is lower than Auger's and therefore more events are seen, but below the correlation threshold.

Significant differences exist between the two data sets. Further data will help to clarify the differences: most important the energy scales of the two experiments; if the Northern sky is significantly different on the nature of its sources; if the Virgo deficit is real and if the AGNs are traces of the real sources and/or the sources themselves.

## 4. SUMMARY

Knowledge of the energy spectrum of the highest energy cosmic rays, their nature and sources will improve significantly in detail as the Pierre Auger Observatory continues to take data in the forthcoming years.

## 5. ACKNOWLEDGMENTS

Work by the author and the USA groups are supported by Department of Energy, Contract No. DE-AC02-07CH11359, and National Science Foundation, Grant No. 0450696. In addition, we acknowledge the support of the science funding agencies of participating countries.